# Characteristics of Flight Delays during Solar Flares

X. H. Xu[1], Y. Wang[1], F. S. Wei[1], X. S. Feng[1], M. H. Bo[2], H.W. Tang[2], D. S. Wang[2], L. Bian[2], B.Y. Wang[1], W. Y. Zhang[1], Y. S. Huang[1], Z. Li[3], J. P. Guo[4], P. B. Zuo[1], C. W. Jiang[1], X.J. Xu[5], Z. L. Zhou[5] and P. Zou[1]

1 Institute of Space Science and Applied Technology, Harbin Institute of Technology, Shenzhen, China
2 Travelsky Mobile Technology Limited, Beijing, China
3 Nanjing University of Information Science and Technology, Nanjing, China
4 Beijing Normal University, Beijing, China
5 State Key Laboratory of Lunar and Planetary Sciences, Macau University of Science and Technology, Macao, China

**Abstract**

   Solar flare is one of the severest solar activities on the sun, and it has many important impacts on the near-earth space. It has been found that flight arrival delays will increase during solar flare. However, the detailed intrinsic mechanism of how solar flares influence the delays is still unknown. Based on 5-year's huge amount of flight data (~$5\times10^6$ records), here we comprehensively analyze the flight departure delays during 57 solar flares. It is found that the averaged flight departure delay time during solar flares increased by 20.68% (7.67 min) compared to those during quiet periods. It is also shown that solar flare related flight delays reveal apparent time and latitude dependencies. Flight delays during dayside solar flares are more serious than those during nightside flares, and the longer (shorter) delays tend to occur in the lower (higher) latitude airport. Further analyses suggest that flight delay time and delay rate would be directly modulated by the solar intensity (soft X-ray flux) and the Solar Zenith Angle. For the first time, these results indicate that the communication interferences caused by solar flares will directly affect flight departure delay time and delay rate. This work also expands our conventional understandings to the impacts of solar flares on human society, and it could also provide us with brand new views to help prevent or cope with flight delays.

## Introduction

   Flares are one of the severest solar activities that release large amounts of electromagnetic energy from the sun (*Hannah et al.*, 2011). Solar flares have great influences on our near-Earth space and they are important topics of space weather research (*Liu et al.*, 2021; *Singh et al.*, 2010). It has been found that solar flares can affect spacecraft operations, short wave communications (*Yasyukevich et al.*, 2018), GPS/GNSS navigation (*Berdermann et al.*, 2018), and air traffic control facilities (*Marque et al.*, 2018). With the development of modern science and technology, the impacts of solar flares on human society have been found to be more and more obviously. During a solar flare, the sun suddenly releases a huge amount of energy (about $10^{28}$ to $10^{32}$ ergs) (*Aschwanden et al.*, 2017; *Emslie et al.*, 2005; *Hudson*, 2011) in few min or hours, and it emits various electromagnetic radiations, most notably X-ray and extreme ultraviolet (*Mendillo et al.*, 2006). Enhanced X-ray and extreme ultraviolet radiations during solar flares can cause a sudden increase of ionization in the Earth's lower ionosphere (D region) (*Quan et al.*, 2021; *Thomson et al.*, 2005). The sudden increase in electron density can lead to a series of disturbances (sudden ionosphere disturbance) which will result in fading or even disruption of shortwave radio signals and interfering with communication and navigation systems. While communication and navigation play an important role in flight operations. In particular, to ensure safe operations, aircraft communication must be available over the entire route (*eCFR*, 2007).

It has been sporadically reported that solar flares could affect the flight operations. In November 2015, the operation of air traffic control radar stations in Sweden were affected by the solar flare associated radio radiations (*Jonas et al.*, 2016; *Marque et al.*, 2018), which disrupted Swedish airspace for nearly an hour and led to flight delays. In September 2017, a series of fairly strong solar flares caused widespread periodic impacts on high-frequency radio users around the world. The French Civil Aviation authorities reported that an aircraft equipped with non-Controller Pilot Data Link Communications equipment lost shortwave radio contact for about 90 min (*Redmon et al.*, 2018). Some authors even found an increasing number of aviation catastrophes on the third day after a X-class flare (*Kalinin and Lukyanova*, 2014).

The impacts of solar flares on the aviation industry have been researched by some international organizations and scholars (*ICAO*, 2018; *UK CAA*, 2020). But their focus is mainly on the safety issues, and there is very little news or reports related to flight delays, which are of great concern to many passengers. Existing researches have clearly shown that flight delays can result in huge economic losses and lead to a series of problems (*Ferguson et al.*, 2013; *Zou and Hansen*, 2014). Therefore, it is very important to delve into the relationships between solar flares and flight delays. Our previous report has preliminarily revealed that the solar flare associated ionospheric disturbances are highly correlated with flight delays (*Wang et al.*, 2022). However, what are the unique characteristics of flight delays during solar flares? What is the underlying mechanism linking flight delays and solar flares? In this study, based on the data of solar flares from GOES satellite and huge amounts (~$5\times10^7$) of flight records, the time and latitude dependency of flight delays during solar flares are revealed and the relevant potential explanations are discussed.

**Data and Methods**

The individual flight data are obtained from the Travelsky Mobile Technology Limited, an affiliated company of the Civil Aviation Administration of China. The flight records include the operational date, flight number, planned/real departure airports, planned/real arrival airports, planned/real departure time and planned/real arrival time from January 1, 2015 to December 31, 2019. These data cover all the commercial flight records in the Guangzhou Baiyun International Airport (IATA: CAN), Shanghai Pudong International Airport (IATA: PVG), Beijing Capital International Airport (IATA: PEK),

Shanghai Hongqiao International Airport (IATA: SHA) and Shenzhen Baoan International Airport (IATA: SZX). These airports are ideal samples for analyses since they are the five largest hub airports in China, which could statistically guarantee the data homogeneity.

The solar flare events are mainly obtained from National Oceanic and Atmospheric Administration (NOAA) (*NOAA-website*) while the unlisted events are determined by using soft X-ray data from GOES satellites. According to the NOAA Space Weather Scales, Solar flares below class M usually do not significantly affect communications, navigation, and positioning. Therefore, only M-class and X-class solar flare events are selected in this analysis. Meanwhile, there is a 24-hour intrinsic period for flight delays (*Wang et al.*, 2022), and to avoid the interference of such 24-hour intrinsic period, the durations of all solar flare events are set to 24 hours. As shown in Fig.1, only independent solar flare events (Case A and Case B) are selected as the research samples, while those solar flares (Case C) interfered by other space weather events (e.g., coronal mass ejections, or solar proton events) are excluded. Some flights data are not used if they are not in an entire day (from 00:00 to 24:00, local time, the green area). Accordingly, the flight data in the red area are selected as be affected by the solar flare events, while the flight data in the blue area are defined as the flight during quiet periods. Finally, 52 solar flare events and 5,295,347 individual flight records from January 1st, 2015 to December 30th, 2019 are selected to use in the analyses.

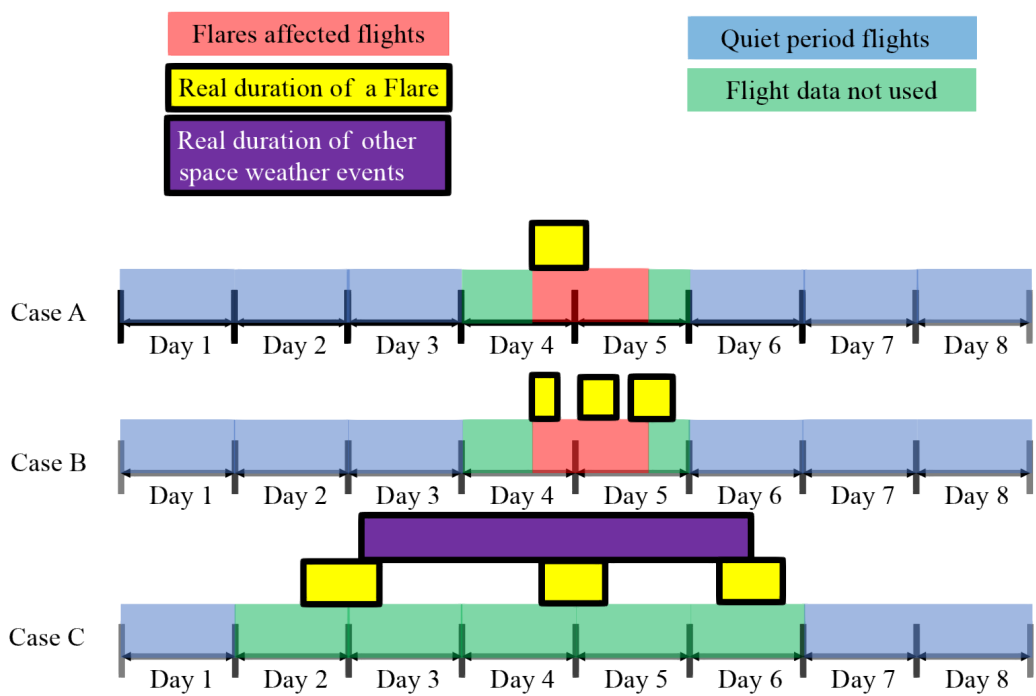

Fig.1. The selection of solar flare affected flights (red) and the flights during quiet periods (blue).

# Results

The effects of solar flares on the Earth depends on many factors, such as the local time (the longitude), the latitude, the flare intensity, and the Solar Zenith Angle (SZA) (*NOAA-website*; *Singh et al.*, 2010; *Yasyukevich et al.*, 2018). Therefore, if the increased flight delay time during solar flares is really caused by solar flares but not statistical coincidences, the delay time should reveal some distinct features related to solar flares. Hence, it also motivates us to investigate the detailed characteristics of flight delays and discuss their internal relationships with solar flares.

First and foremost, to have a thorough understanding to the impacts of solar flares on flight departure delays, the distributions of flight departure delays based on all individual flight data are calculated and shown in Fig. 2(a). It is found that compared to quiet periods, the distribution of flight delays during solar flares shifts to the right as a whole, which means an overall larger delay time during solar flares. The average delay time during solar flares is 44.75 min (median: 25 min), which is 7.67 min (median: 5 min) longer than those during quiet periods. These results suggests that flight departure delays will increase during solar flares and such scenario is consistent with our previous studies to the arrival delays (*Wang et al.*, 2022).

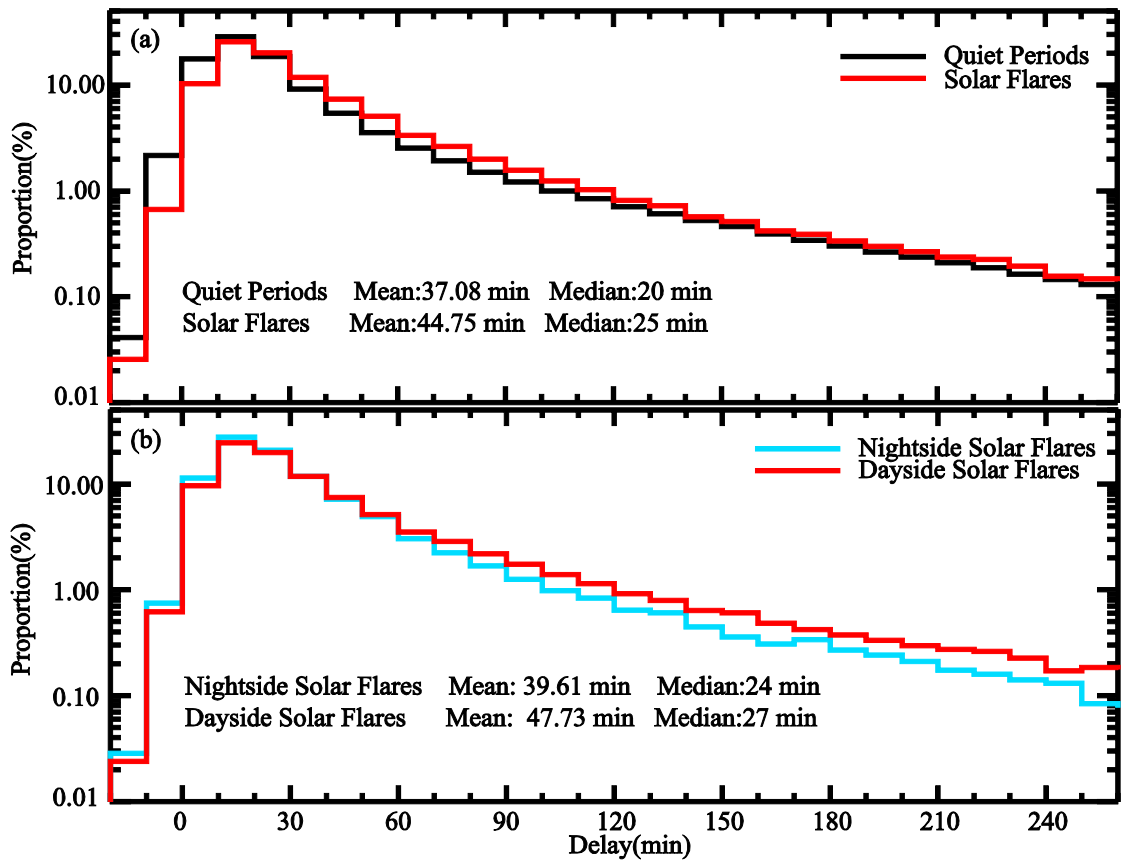

Fig.2 (a)Distributions of flight departure delay time during solar flares (red) and quiet periods (black) averaged over five airports. (b) Distributions of flight departure delay time during dayside solar flares (red) and nightside solar flares (blue) averaged over five airports.

In addition, local time is one of the most obvious factors that affects the Earth's response to solar flares (*Zhang and Xiao*, 2000). Therefore, if the flight delays are caused by solar flares, there should be significant differences between the dayside and nightside. Theoretically, the effects of solar flare events on the earth is mainly concentrated in the subpoint and varies smoothly with the cosine of SZA (*NOAA-DRAP*; *Qian et al.*, 2019). It can be inferred that the HF radio absorption will be prominent in the dayside ionosphere, while hardly any absorption will occur in the nightside ionosphere. Therefore, during solar flares, communications and navigation of flights on the dayside could be severely affected, while those on the nightside would have little impacts. Accordingly, we comparatively analyzed the characteristics of solar flare related flights during dayside and nightside. If the start time of a solar flare is between 06:00 and 18:00 (local time), it is defined as the dayside solar flare, and otherwise, defined as nightside solar flare. Fig. 2(b) shows the distribution of flight departure delays during 33 dayside solar flares and 19 nightside solar flares averaged over five airports. Two obvious features can be seen from the figure. First of all, regardless of nightside and dayside, the averaged flight delay time during solar flares is larger than those during quiet periods. The duration of flares can last several hours and the aviation activities also need time to response, and such circumstances may partially account for the increased delay time during nightside solar flares. In addition, the distribution of the delay time during dayside solar flares also rises on the right side compared to those during nightside solar flares. These results confirm with our expectations that the flight delays are more serious during dayside solar flare, and they also suggest that flare affected flight delays reveal obvious time dependency.

Last but not least, latitude is also an important factor that controls the effects of solar flares on Earth (*NOAA-DRAP*; *Zhang et al.*, 2002). The investigated airports are all located around the similar longitudes but well separated in latitudes, which provides an ideal condition to reveal the latitude effect. According to their locations, the five airports are classified as three categories, the higher latitude airport: PEK (40.07°N, 116.58°E), the middle latitude airport: SHA (31.19°N, 121.33°E) and PVG (31.15°N, 121.80°E), and the lower latitude airport: CAN (23.38°N, 113.30°E) and SZX (22.64°N, 113.82°E). In general, lower latitudes will be affected more significantly than the higher latitudes (above the tropics) (*Pettit et al.*, 2018; *Qian et al.*, 2011). The flare associated interferences in communication and navigation will be more severely at lower latitude airport than those at higher latitude airport. Therefore, it is expected that the lower (higher) latitude airport will reveal the longer (shorter) flight delays.

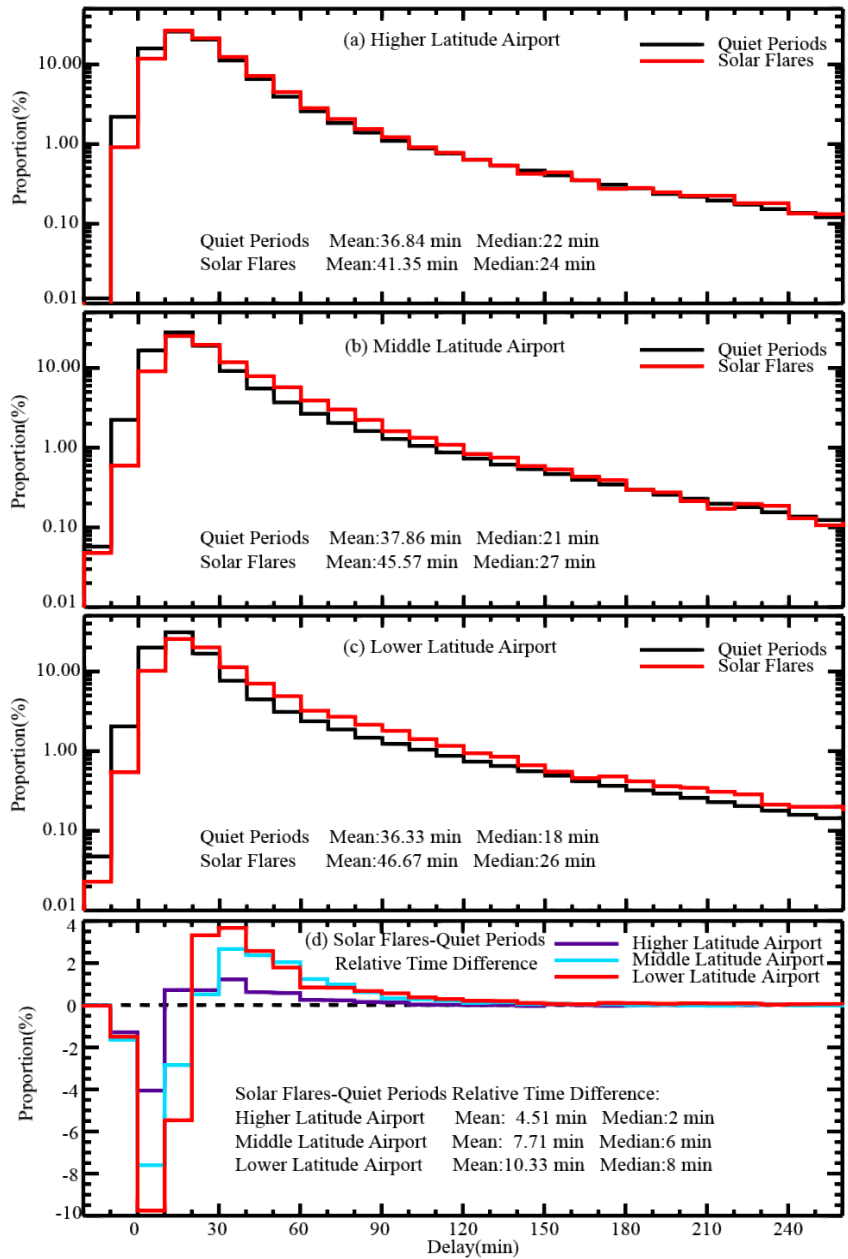

Fig.3 Distributions of flight departure delay time during solar flares (red) and quiet periods s (black) at higher (a), middle (b), and lower (c) latitudes airports respectively. (d) Distributions of the relative difference of delay time between those during solar flares and quiet periods (higher latitudes airport: purple, middle latitudes airport: blue, lower latitudes airport: red).

Fig. 3 (a), (b) and (c) show the departure delay distribution of three types of airports during solar flares and quiet periods respectively. One can see that the flight delay time during solar flares is 46.67 min, 45.57 min, and 41.35 min (median: 26 min, 27 min, 24 min) respectively from lower latitude to higher latitude airports. Meanwhile the average delays during quiet periods are nearly unchanged (~37 min). Fig. 5 (d) shows the relative difference of delay time between those during solar flares and quiet periods, which could also be regarded as the net increase in the flare caused delay time.

The latitude effect of flight delays behaves very clearly that the lower latitude airport always has higher flight delays. It is worth noting that the net increases in average delay time show a smooth trend of change with latitude (10.33 min near ~23°N, 7.71 min near ~31°N and 4.51 min near ~40°N), and the change rate is ~0.35 min/degree. These results indicate that flight delay time affected by solar flares has distinct latitude effect as expected.

## Discussions and Conclusions

Previous studies has indicated that flight arrival delays will be affected by solar flares (*Wang et al.*, 2022). It is found in this analyses that the solar flare associated flight departure delays are also similar to the arrival delays. In addition, the definite time and latitude dependency of the delay time further imply that there should exist a causal relationship between solar flares and flight delays but not just coincidences. To investigate the underlying mechanism linking flight delays and solar flares, we would like to focus on the flare intensity and the SZA.

The intensity of the solar flares, represented by the soft X-ray flux, is an important index of solar flares. It is known that the flare associated effects on Earth would be more significant if the flux intensity is more intense. Besides, the impacts of solar flares on Earth are not only related to the flux of X-ray flux, but also related to the SZA. The effect of SZA has been taken into account by many scholars when researching the impacts of solar flares on our human activities (*Wan et al.*, 2005; *Zhang and Xiao*, 2003). In particular, the flare associated HF radio absorption can be estimated by using the DRAP model (*NOAA-DRAP*) . The Highest Affected Frequency (HAF) is introduced to calculate the highest frequency affected by absorption of 1 dB due to solar X-ray flux (*Schumer*, 2009).

$$HAF = \left[10\log\left(flux\right) + 65\right](\cos\chi)^{0.75}$$

Where the *flux* represents the soft X-ray flux and χ is the SZA. The HAF is an empirical formula which can be used to simply evaluate the interferences of HF communication caused by solar flares. Obviously, the higher X-ray flux and the smaller SZX will lead to more significant consequences.

Fig.4 shows the flight departure delay time and delay rate as a function of HAF during solar flares averaged over five airports. It is noted that the delay time and delay rate reveal a rough increase with the HAF. Gentle increases or fluctuations in flight delays are found when HAF is relatively small. However, the most notable point is that the growth rate of both the delay time and the delay rate boost violently when HAF above 5Mhz. This phenomenon once again confirms the delay effects caused by solar flares and it also implies that the flare associated interferences in HF communication should be the most probable reason accounting for the increased flight delays.

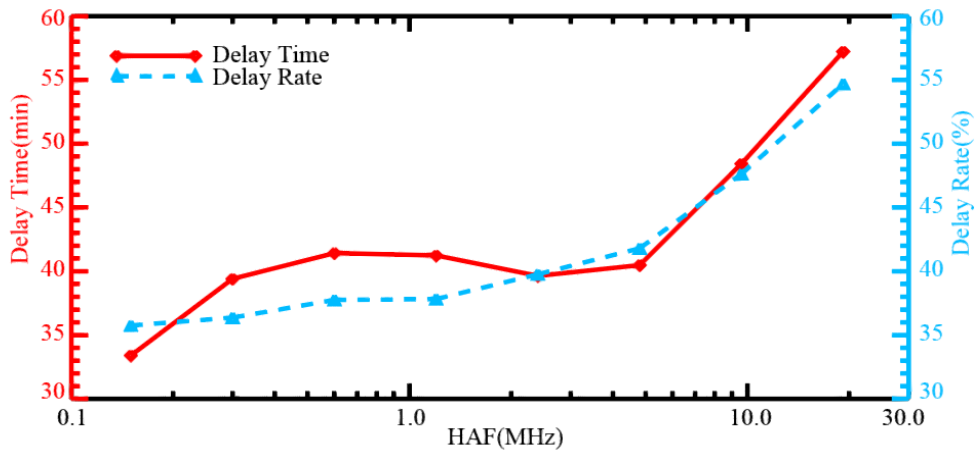

Fig. 4 The flight departure delay time (red) and delay rate (blue) as a function of HAF during solar flares averaged over five airports.

In summary, this work analyses the unique characteristics of flight delays during solar flares and try to reveal the internal relationships between flight delays and solar flares. The averaged flight departure delay time in all airports during solar flares are found to be 7.67 min (20.68%) longer than those during solar flares. The flight delay time also show obvious time dependency that the longer (shorter) flight delays are always found during dayside (nightside) flares. Meanwhile, flight delays also exhibit latitude dependency, and it is found that the lower (higher) latitude airport has longer (shorter) flight delays. We further explore the conclusive factors affecting the flight delay during solar flares by analyzing the relationship between HAF and flight delays. It is indicated that both the soft X-ray flux and the SZA play important roles in flight delays. These results demonstrate that solar flare would account for the increased flight delays, and it is suggested that the communication interferences caused by solar flares would significantly affect flight delays. It is widely known that solar flares have many important impacts on human life, while this work expands our conventional understandings to the solar flares, and it could also provide us with brand new views to help improve the flight delay predications.

## Acknowledgements

This work is jointly supported by the National Natural Science Foundation of China (41731067 and 42174199), Guangdong Basic and Applied Basic Research Foundation (2021A1515012581), and the Shenzhen Technology Project (GXWD20201230155427003-20200804210238001).